\documentclass[aps,prd,superscriptaddress, nofootinbib,showkeys,showpacs,preprint,longbibliography,amsmath]{revtex4-1}
\usepackage{latexsym,verbatim,amsfonts,graphicx,subfigure}

\usepackage{graphicx}
\usepackage{dcolumn}
\usepackage{bm}
\usepackage{rotating}
\usepackage{slashed}

\begin{document}
\title{Semiclassical strings and Non-Abelian T-duality}
\author{S.~Zacar\'{\i}as}
 \email{szacarias@fisica.ugto.mx}
\affiliation{Department of Physics, Swansea University, Singleton Park, Swansea SA2 8PP, United Kingdom.}

\affiliation{Departamento de F\'{\i}sica, Divisi\'{o}n de Ciencias e Ingenier\'{\i}as, Campus Le\'{o}n, Universidad de Guanajuato, Loma del Bosque No. 103 Col. Lomas del Campestre, C.P. 37150, Le\'{o}n, Guanajuato, M\'{e}xico.}

\begin{abstract}
We study semiclassical strings in the Klebanov-Witten and in the Non-abelian T-dual Klebanov-Witten backgrounds.
We show that both backgrounds share a subsector of equivalent states up to conditions on the T-dual coordinates which also impose a breaking
of the R-symmetry by a discrete group. We also analyze string configurations where the strings are stretched along the T-dual coordinates.
This semiclassical analysis predicts the existence of chiral primary operators for the dual (strongly coupled) superconformal field theory whose anomalous
dimension depends on the T-dual coordinates. We briefly discuss the Penrose limit of the dualised background.
\end{abstract}

\maketitle

\section{Introduction}
Dualities in string theory have played an important role in the developments of the theory in the last two decades
providing new insigths into the non-perturbative properties of the theory. For instance, we have learnt how to relate 
string theories in different regimes (S-duality) and compactifications  (T-duality). 
Moreover, the study of the theory at a non-perturbative level led to the discovery of higher dimensional objects, D-branes. 
The dual description of such objects as hyperplanes whose dynamics is described by open strings attached to them \cite{Polchinski:1995mt} and as source of closed strings 
\cite{Gubser:1997yh,Gubser:1997se} led to the open/closed string duality. 
These ideas prompted Maldacena \cite{Maldacena:1997re}  to propose the celebrated gauge/string duality. 
The first and best understood example is the equivalence between string theory on $AdS_5\times S^5$ and an $\mathcal{N}=4$ supersymmetric
$SU(N)$ gauge theory. This duality implies that there is a unique prescription relating physical quantities in the string
and field theory sides. In particular it implies the relation between quantum numbers 
like AdS energy as function of the tension, angular momenta and spin of a string with the dimension of corresponding field theory operators,
 being this quantities computed where either string theory and field theory can be analyzed quantitatively.
The precise checking of the correspondence, apart from the special case of protected (BPS) quantities, was difficult to establish until the notable
discovery that for a particular class of operators the comparison of both sides is possible \cite{Berenstein:2002jq}. This corresponds to 
operators with large quantum numbers. This idea was clarified and extended in \cite{Gubser:2002tv} leading to semiclassical limits
where the duality can be extended. 
More examples of this gauge/string duality have been analyzed generalizing this idea to models of a reduced number of supersymmetries either preserving or 
not conformal invariance (see \cite{Aharony:2002up,Bigazzi:2003ui}  for a review). Also the same semiclassical analysis has been applied to these examples
giving strong hints on how the emergent field theories can be investigated.

Moreover, the old idea of generalizing T-duality to include non-Abelian isometry groups \cite{Ossa:1992vc}, has been recently explored as a
solution generating technique for supergravity \cite{Sfetsos:2010uq, Lozano:2011kb, Itsios:2012dc, Lozano:2012au, Jeong:2013jfc,Itsios:2012zv, Nunez}.  This idea has been applied to $\mathcal{N}=1$ 
backgrounds, either conformal or non conformal, containing an $SU(2)$ isometry group whose gauge theory dual (strongly coupled) is well understood \cite{Itsios:2012zv, Nunez}.
 A common feature of these backgrounds is that they retain supersymmetry \cite{Nunez,Barranco:2013fza}. It is in this spirit that the use of 
non-Abelian T-duality provide a good tool to construct supergravity duals of strongly coupled field theories and at the same time to overcome guessing new solutions of type II 
supergravity. For instance, applying this procedure along the $SU(2)$ isometry of the internal space of the Klebanov-Witten (KW)
background results in a solution of type IIA supergravity which retains the $AdS_5$ factor and whose lift to M-theory corresponds to geometries 
obtained in \cite{Bah:2012dg}  from wrapping M5 branes on $S^2$. 
The solutions constructed using non-Abelian T-duality present a host of interesting phenomena.
Recent developments studying the geometric picture of the backgrounds obtained have revealed some
effects of this phenomena in the dual field theory \cite{Barranco:2013fza}. 
For instance, using the language of G-structures, it was shown for the baryonic branch of the  Klebanov-Strassler field theory that the non-Abelian T-dual solution is such that 
 the phenomena of confining and symmetry breaking are  encoded in changes on the character of these structures \cite{Gaillard:2013vsa}.
 There are, however, many points that deserve study towards the gauge/string realisation of this T-dual solutions.

Here we shall investigate rotating and spinning closed strings in the KW and the non-Abelian  T-dual KW backgrounds
with two components of angular momentum in the internal space. 
We consider configurations which exhaust the number of maximal translational isometries that
we can have before and after the non-Abelian T-duality is implemented. Among these isometries there is a $U(1)_R$ factor associated to shifts along the
coordinate $\psi$ which behaves as the R-symmetry before and after the non-Abelian T-duality \cite{Gaillard:2013vsa}. 
This will play a crucial role in our discussion. We remark that the semiclassical analysis of strings in the KW background has been already studied in the 
literature (see, e.g.  \cite{Tseytlin:2002ny,Schvellinger:2003vz,Benvenuti:2005cz,Kim:2003vn,Pons:2003ci}). 
However, the solution considered here is of particular interest for the comparison with strings moving in the T-dual background
generalizing certain previous known cases which will reduce to these by taking different limits.

We first study the behavior of conserved quantities in the KW background 
in certain interesting limits to recover the expected features of energy depending on spin and angular momenta.  Our goal will be to understand 
how some of these quantities change or not under non-Abelian T-duality.
We then study these string configurations in the T-dual background.
We show that there is an equivalence between states in both backgrounds which implies conditions on the T-dual coordinates.
This shows that, in spite of leaving untouched some properties (isometries) of the original background, non-Abelian T-duality acts in a non trivial way
by mixing the R-charge isometric direction with the T-dual coordinates as a consequence of a gauge fixing choice in the procedure. We also study configurations which 
involve closed strings stretched along the T-dual coordinates. We shall see that for finite energy configurations certain 
solutions are not allowed. Note that for strings in the non-Abelian T-dual background we will have non-zero B field on the worldsheet. 
However, for the cases we are interested here this will not introduce corrections.

In Section \ref{sec1} we briefly review the basic tools to implement the non-Abelian T-duality transformations for backgrounds that support an $SU(2)$ isometry.
In Section \ref{sec2} we apply this technique to the KW background.
In Section \ref{sec3.1} we study semiclassical closed strings in the KW background 
to analyze the relationship between conserved quantities that arise like energy, spin and angular momenta. We recover the expected features
for large and short strings. We then study closed string configurations in the non-Abelian T-dual KW background to analyze how the states behave under non-Abelian T-duality. 
We also study strings which are stretched along the T-dual coordinates.
In section \ref{sec5} we briefly discuss the Penrose limit for the T-dual background. 
Section \ref{sec6} is devoted to some concluding remarks.


\section{Non Abelian T-duality}\label{sec1}
In this section we briefly review the non-Abelian T-duality technique, a comprehensive treatment may be found in \cite{Nunez}.

T-duality is a symmetry transformation that relates different string backgrounds possessing some isometries.
This is a full symmetry of string (genus) perturbation theory. In its simple form, 
it is the statement that strings do not make a distinction between large and small compact spaces on which they propagate. 
The T-duality rules, known as Buscher rules, relate two generically different but dynamically equivalent sets of background fields.

In the path integral formulation of T-duality a central role is played by the isometry group $U(1)$ of the background one chooses to dualise.
By gauging this $U(1)$ isometry, enforcing a flat connection for the corresponding gauge field with a Lagrange multiplier and integrating out
this Lagrange multiplier one produces the T-dual sigma model.
When the isometry group is not Abelian one may follow a similar path to naturally arrive at the notion of non-Abelian T-duality  \cite{Ossa:1992vc}. However, the similarities stop here. 
While in the Abelian case the transformation is invertible in the non-Abelian counterpart the isometry is, at least partially, destroyed.
However this lost isometry may be recovered as a non-local symmetry in the sigma model and the corresponding sigma models are canonically equivalent \cite{Curtright:1994be}. 
A second subtle difference is that due to global complications,
it is thought that non-Abelian T-duality is not a full symmetry of string perturbation theory. However, it remains valid 
as a solution generating symmetry of supergravity. 

Let us consider a bosonic string sigma model in a NS background that supports an $SU(2)$ 
isometry, such that the background fields can be expressed in terms of left-invariant 
Maurer-Cartan forms $L^{i}=-i\textrm{Tr}(g^{-1}dg)$. That is to say the target space metric can be written as
\begin{gather}
ds^2=G_{\mu\nu}(x)dx^{\mu}dx^{\nu}+2G_{\mu i}(x)dx^{\mu}L^{i}+g_{ij}(x)L^{i}L^{j},
\label{metricdec}
\end{gather}
where $\mu,\nu=1,2,...7$, with corresponding expressions for the NS two form, $B$, and dilaton, $\Phi$. 
Hence, all the coordinate dependence on the $SU(2)$ Euler angles $\theta,\phi, \psi$ is contained in the Maurer-Cartan forms
whilst the remaining data can all be dependent on the spectator fields $x^{\mu}$.

The non-linear sigma model is
\begin{gather}
S=\int d^{2}\sigma\left(Q_{\mu\nu}\partial_{+}x^{\mu}\partial_{-}x^{\nu}+Q_{\mu i}\partial_{+}x^{\mu}L_{-}^{i}+Q_{i\mu}L_{+}^{i}\partial_{-}x^{\mu} +E_{ij}L_+^{i}L_{-}^{j}\right),
\label{sigmam}
\end{gather}
where
\begin{gather}
Q_{\mu\nu}=G_{\mu\nu}+B_{\mu\nu}, \quad Q_{\mu i}=G_{\mu i}+B_{\mu i}, \quad Q_{i\mu}=G_{i\mu}+B_{i\mu}, \quad E_{ij}=g_{ij}+b_{ij},
\label{def}
\end{gather}
where $L_{\pm}^{i}$ denotes the pull back of the left-invariant forms to the worldsheet. The first step in performing the non-Abelian procedure 
is to gauge the isometry by replacing derivatives by covariant derivatives in the Maurer-Cartan forms. Then we add the Lagrange multiplier term 
$-i\textrm{Tr}(vF_{+-})$ to enforce a flat connection. After integrating this Lagrange multiplier term by parts, 
one can solve for the gauge fields to obtain the T-dual model that still depends on $\theta, \phi,\psi, v_i$ and the spectators. We must gauge fix this redundancy.  
 We shall fix the three Euler angles to zero, i.e. $g=\mathbb{I}$. Nonetheless, one can use a more general gauge fixing in order to 
make manifest some residual symmetries.

We obtain the Lagrangian
\begin{gather}
S=\int d^{2}\sigma\left(Q_{\mu\nu}\partial_{+}x^{\mu}\partial_{-}x^{\nu}+(\partial_{+}v_{i}+\partial_{+}x^{\mu}Q_{\mu i})(M_{ij})^{-1}(\partial_{-}v_{j}-Q_{j\mu}\partial_{-}x^{\mu})\right),
\label{tduals}
\end{gather}
where $M_{ij}=E_{ij}+f_{ij}^{k}v_{k}$.
 Thus, it is clear from the above Lagrangian that the non-Abelian background fields will
 be determined by the inverse of $M_{ij}$.
As in the Abelian case the dilaton receives a contribution as a consequence of the above manipulations in the path integral.

For type II supergravity backgrounds it is also necessary to know how the RR fluxes transform under 
non-Abelian T-duality. After dualisation, left and right movers couple to different sets of frame fields.
However, since these frame fields define the same geometry they must be related by a Lorentz transformation 
$\Lambda$. In the present case, one can ascertain that $\textrm{det}\Lambda=-1$ and consequently the dualisation
maps between type IIB and type IIA theories.
This Lorentz transformation also induces an action on spinors given by a matrix $\Omega$ by requiring that
\begin{gather}
\Omega^{-1}\Gamma^{a}\Omega=\Lambda_{b}^{a}\Gamma^{b}.
\end{gather}
The non-Abelian T-dual fluxes are obtained by right multiplication with the above $\Omega$ matrix, namely
\begin{gather}
e^{\Phi_{\textrm{IIA}}}{\slashed F}_{\textrm{IIA}}=e^{\Phi_{\textrm{IIB}}}{\slashed F}_{\textrm{IIB}}\cdot \Omega^{-1}.
\end{gather}
where $F_{{\textrm{IIA}}/{\textrm{IIB}}}$ are  RR polyforms and the slash notation indicates that we have converted these 
polyforms to bispinors by contraction with gamma matrices.

Another feature of some type II supergravity backgrounds is that they enjoy certain amount of supersymmetries. 
Whether (and how much) supersymmetry is preserved depends on how the Killing vectors about which we dualise act on the supersymmetry.
The Lie derivative on spinors is given by \cite{FigueroaO'Farrill:1999va}
\begin{gather}
\mathcal{L}_{k}\epsilon=k^{\mu}D_{\mu}\epsilon +\frac{1}{4}\nabla_{\mu}k_{\nu}\gamma^{\mu\nu}\epsilon.
\label{derivative}
\end{gather}
If, when acting on the Killing spinors of the initial geometry, this vanishes automatically for all the Killing vectors generating the action of 
the group of isometries, then we anticipate that supersymmetry will be preserved  in the dual background. If on the other hand this vanishes only for 
some projected subset of Killing spinors then, we expect only a corresponding projected amount of supersymmetries to be preserved.
In the case we are interested here, one can show that (\ref{derivative}) vanishes identically along the $SU(2)$ isometries \cite{Itsios:2012zv,Nunez}. This corresponds to the statement that in the dual field theory the supersymmetry 
is not charged under this $SU(2)$ flavour symmetries.  

Parallel to this discussion is the fact that the $U(1)_R$ symmetry commutes with $SU(2)$ and hence one expects the corresponding isometry to be preserved under dualisation.


\section{Dualisation of the Klebanov-Witten background}\label{sec2}
The stack of D3 branes at the tip of the conifold was studied within the AdS/CFT correspondence in \cite{Klebanov:1998hh}.
The gauge theory describing the low energy dynamics of the branes is an $\mathcal{N}=1$ superconformal field theory with 
product gauge group $SU(N)\times SU(N)$. There are two sets of bifundamental matter fields; $A_i$ in the $(N,\overline{N})$ 
and $B^{m}$ in the $(\overline{N},N)$. The indices $i$ and $m$ correspond to two sets of $SU(2)$ global symmetries. 
The super potential for the matter fields is given by
\begin{gather}
W=\frac{\lambda}{2}\epsilon^{ij}\epsilon_{mn}Tr(A_{i}B^{m}A_{j}B^{n}).
\end{gather}

This gauge theory is dual to string theory on $AdS_{5}\times T^{1,1}$ with $N$ units of RR flux on $T^{1,1}$.
Here $T^{1,1}$ is a $U(1)$ bundle over $S^2\times S^2$ with Einstein metric satisfying $R_{ij}=4g_{ij}$. The metric and the 5-form self-dual flux, are given by
\begin{align}
ds^{2}=&ds_{AdS_{5}}^2+ds^{2}_{T^{1,1}}\label{background},\\
F_{(5)}=&\frac{4}{g_sL}(vol(AdS_5)-L^5vol(T^{1,1}))\label{f5},
\end{align}
where 
\begin{align}
ds_{AdS_{5}}^2=&L^2\left(-\textrm{cosh}^2\rho dt^2+d\rho^2+\textrm{sinh}^2\rho (\textrm{cos}^2\beta d\hat{\phi}+d\beta^2+\textrm{sin}^2\beta d\phi^2 )  \right),\label{ads}\\
ds^{2}_{T^{1,1}}=&L^2\left(\frac{1}{6}(d\theta_1^2+\textrm{sin}^2\theta_1 d\phi_1^2)+\frac{1}{6}(d\theta_2^2+\textrm{sin}^2\theta_2 d\phi_2^2)\right.\nonumber \\
&\left.+\frac{1}{9}(d\psi +\textrm{cos}\theta_1 d\phi_1+\textrm{cos}\theta_2 d\phi_2)^2\right),\label{t11}
\end{align}
where $L^2=\sqrt{\lambda_t}\alpha'\sqrt{27/4}$ is the curvature radius of $T^{1,1}$ and $\lambda_t=4\pi g_{s}N$ the 't Hooft coupling.
We have chosen the coordinates $(\theta_i,\phi_i)$, $i=1,2.$ to parametrize the two $S^2$ in $T^{1,1}$ whilst the angle $\psi$ has period $4\pi$.
The group of isometries for this metric is  $SU(2)_1\times SU(2)_2\times U(1)_R$ and is identified with the $SU(2)_1\times SU(2)_2$ global symmetry and 
$U(1)_{R}$ R-symmetry of the dual superconformal field theory.

We will work with the following frame fields for this background
\begin{gather}
e^t=L\cosh\rho dt,\quad e^{\beta}=L\sinh\rho d\beta,\quad e^{\phi}=L\sinh\rho\sin\beta d\phi,\nonumber\\
e^{\hat{\phi}}=L\sinh\rho\cos\beta d\hat{\phi},\quad e^{\rho}=Ld\rho,\\
e^{\hat{1},\hat{2}}=\lambda_1\sigma_{\hat{1},\hat{2}},\quad e^{1,2}=\lambda_2\sigma_{1,2},\quad \sigma_3=\lambda(d\psi+\cos\theta_1d\phi_1),
\end{gather} 
where $\sigma_{\hat{1}}=\sin\theta_1 d\phi_1$ and $\sigma_{\hat{2}}=d\theta_1$ are 
the frame fields for the $S_1^2$ and $\lambda^2_1=\lambda^2_2=\frac{1}{6}$, $\lambda^2=\frac{1}{9}$.
We have also introduced the left-invariant one forms parametrised by Euler angles as follows
\begin{gather}
\sigma_1=\cos\psi\sin\theta_2 d\phi_2-\sin\psi d\theta_2,\quad \sigma_2=\sin\psi\sin\theta_2 d\phi_2+\cos\psi d\theta_2,\nonumber\\
\sigma_3=d\psi+\cos\theta_2 d\phi_2.
\end{gather}
We dualise with respect to the $SU(2)_2$ global symmetry defined by the above $\sigma'_i$s
and we partially gauge fix $SU(2)_2$ by retaining the $\psi$ coordinate related to the R symmetry. The result of the dualisation 
 was presented in \cite{Itsios:2012zv, Nunez}. The result is an $\mathcal{N}=1$
supersymmetric solution of type IIA with NS and RR fields given by\footnote{We have set $L=1$ which may be restored by appropriate rescalings.}
 \begin{gather}
  ds^2_{\textrm{dual}}=ds^{2}_{AdS_5}+d\hat{s}^{2}_{T^{1,1}}\label{penrose}\\
    e^{-2\hat{\Phi}}=\frac{8}{g_s^2}\Delta\label{dil}\\
  \hat{B}=-\frac{\lambda^2}{\Delta}\left[x_1x_2dx_1+(x_2^2+\lambda_2^4)dx_2\right]\wedge\sigma_{\hat{3}}\label{bfield}\\
\hat{F}_4=\frac{8\sqrt{2}}{g_s}\lambda_{1}^2\lambda_2^2\lambda\frac{x_1}{\Delta}\sigma_{\hat{1}}\wedge \sigma_{\hat{2}}\wedge \sigma_{\hat{3}}\wedge(\lambda_2^2x_1dx_2-\lambda^2x_2dx_1),
  \quad\hat{F}_{2}=\frac{8\sqrt{2}}{g_s}\lambda_1^4\lambda\sigma_{\hat{1}}\wedge\sigma_{\hat{2}},
\label{dualfluxes}
  \end{gather}
  where $\sigma_{\hat{3}}=d\psi+\cos\theta_1 d\phi_1$ and the metric on the dualised background is explicitly
  \begin{align}
 d\hat{s}^{2}_{T^{1,1}}=&\lambda_{1}^{2}\left(\sigma^{2}_{\hat{1}} +\sigma^{2}_{\hat{2}}\right)+\frac{\lambda^{2}_{2}\lambda^{2}}{\Delta}x_{1}^{2}\sigma^2_{\hat{3}}\nonumber\\
 &  +\frac{1}{\Delta}\left(\left(x_{1}^{2}+\lambda^2\lambda_{2}^{2} \right) dx_{1}^2+\left(x_{2}^{2}+\lambda_{2}^{4} \right)  dx_{2}^2+2x_1x_2dx_1dx_2\right),
  \label{metricdual}
\end{align}
where 
\begin{gather}
\Delta=\lambda_2^2x_1^2+\lambda^2(x_2^2+\lambda_1^4).
\label{delta}
\end{gather}
The metric in eq. (\ref{metricdual}) has evidently an $SU(2)_1\times U(1)_{R}$ isometry and for a fixed value of $(x_1,x_2)$ the remaining directions give a squashed three sphere.

\section{Classical solutions for rotating strings}\label{sec3.1}
In this section we consider a rotating and spinning closed string in the KW background (\ref{background})-(\ref{f5}) and in the dualised
KW background (\ref{penrose})-(\ref{dualfluxes}) with two components of angular momentum. Such configurations will generate a subgroup
of isometries enjoyed by both backgrounds. We study configurations by fixing two of the three coordinates of $SU(2)_2$ in (\ref{t11}) and the T-dual 
coordinates in (\ref{metricdual}). We also consider closed strings which
are stretched along the T-dual coordinates. Since the aim of this paper is to study the effect of non-Abelian T-duality in this semiclassical configurations
and because the $AdS_5$ comes as an spectator sector, we will keep fixed its string configuration throughout this section. 
We follow closely the procedure outlined in \cite{Russo:2002sr}, in the way used to relate conserved quantities in certain interesting
limits as well as the $AdS_5$ configuration considered there, although the discussion here
is independent of it. 

\subsection{Classical solution for rotating strings in the Klebanov-Witten Background} 

The classical solution for a closed string spinning along the $\phi$ direction of $S^3$ in $AdS_5$ and rotating
simultaneously along $\phi_1$ and the R-charge direction, $\psi$,  within $T^{1,1}$, which is stretched along the radial direction 
and  in the angular coordinate  $\theta_1$ of $S_1^2$ in $T^{1,1}$, can be parametrized by the ansatz
\begin{gather}
t=\kappa \tau,\qquad \phi=\omega\tau ,\qquad \qquad \hat{\phi}=0,\qquad \beta=\frac{\pi}{2},\nonumber\\
\rho=\rho(\sigma)=\rho(\sigma+2\pi),\qquad \theta_1=\theta_1(\sigma)=\theta_1(\sigma+2\pi), \label{solution}\\ 
\theta_2=\textrm{fixed},\qquad  \phi_2=\textrm{fixed},\qquad \phi_1=v_{\phi_{1}}\tau,\qquad \psi=v_{\psi}\tau. \nonumber
\end{gather}
Besides the isometries of $AdS_5$, this configuration generates a subgroup of isometries $U(1)_{1}\times U(1)_R$ of (\ref{t11}) which corresponds to the translational isometries
in $\phi_1$ and $\psi$ that one can relate with charges of operators in the dual superconformal field theory. We can generalize this configuration by considering also shifts in $\phi_2$ which enlarge the
isometry subgroup to $U(1)_{1}\times U(1)_2\times U(1)_R$ but, as we have seen, this shift symmetry is no longer present in the T-dual background.
The equations of motion become
 \begin{align}
 \rho''=&(\omega^2-\kappa^2)\cosh\rho\sinh\rho=0,\label{eqrho}\\
 \theta''_1=&-\frac{v_{\phi_1}^2}{3}\sin 2\theta_1+\frac{2}{3}v_{\phi_1}v_{\psi}\sin\theta_1\label{eqtheta}, 
 \end{align}
where prime denotes derivative with respect to $\sigma$. The solution for 
$\rho$ was computed in \cite{Russo:2002sr} while the one for $\theta_1$ follows from the conformal constraints. The general solutions are
 \begin{align}
 (\rho')^2=&\kappa^2\textrm{cos}^2\alpha_0-(\omega^2-\kappa^2)\textrm{sinh}^2\rho,
 \label{solrho}\\
 (\theta'_{1})^2=&\frac{4}{3}v_{\phi_1}^2\textrm{sin}^4\frac{\theta_1}{2}+(\frac{8v_{\phi_1}v_{\psi}}{3}-\frac{4}{3}v_{\phi_1}^2)\textrm{sin}^2\frac{\theta_1}{2}+6\kappa^2\textrm{sin}^{2}\alpha_0-\frac{2}{3}v_{\phi_1,\psi}^2,
 \label{soltheta}
 \end{align}
 where $v_{\phi_1,\psi}=v_{\phi_1}+v_{\psi}$ and $\alpha_0$ is an integration constant. The parametrization of this integration constant will be determining the size of the closed string
which is stretched from $\rho=0$ up to some $\rho_m$.
 
 Let us define three parameters 
 \begin{align}
 a=\frac{v_{\phi_1}^{2}/3}{6\kappa^2\textrm{sin}^2\alpha_0-\frac{2}{3}v_{\phi_1,\psi}^2},\qquad b=\frac{2v_{\phi_1}v_{\psi}/3}{6\kappa^2\textrm{sin}^2\alpha_0-\frac{2}{3}v_{\phi_1,\psi}^2},\qquad c=\frac{\omega^2-\kappa^2}{\kappa^2\textrm{cos}^2\alpha_0}.
 \label{parameters}
 \end{align}
We are interested in finite energy configurations. Such configurations are allowed if $\omega^2>\kappa^2$, consequently $c>0$. We also choose $b>0$ which implies $a>1$.
 
 For the single-fold string \cite{Gubser:2002tv} the interval $0\leq\sigma\leq 2\pi$ is split into four segments; $\theta_1$ and $\rho$ start at $\rho(\sigma)=\theta_1(\sigma)=0$ at $\sigma=0$
  and increase up to a maximal value $\rho_m$ and $\theta_{1m}$ where $\rho'$ and $\theta'_1$ vanish, which in the present case is at $\sigma=\frac{\pi}{2}$. The periodicity condition thus implies an extra relation between the parameters, namely 
  \begin{align}
  \tan^2\alpha_0=\frac{g^2{}_2F^2_{1}\left(\frac{1}{2},\frac{1}{2},1 ;k^2\right)(1+2a(1+\frac{b}{2a})^2)c}{6~{}_2F^2_{1}\left(\frac{1}{2},\frac{1}{2},1 ;-\frac{1}{c}\right)},
  \label{alphazero}
 \end{align}
 where
  \begin{align}
  g=&\frac{2\sqrt{2}}{\sqrt{-1+2a-2b+2\sqrt{(a-b)^2-a}}},\nonumber\\
 k^2=&\left( \frac{-1+2a-2b-2\sqrt{(a-b)^2-a}}{-1+2a-2b+2\sqrt{(a-b)^2-a}} \right).
 \end{align}
   The energy and spin for this soliton are \cite{Russo:2002sr}
  \begin{align}
 E(a,b,c)=&\frac{\sqrt{\lambda_t}}{\sqrt{c}~\textrm{cos}\alpha_0}~{}_2F_1\left(-\frac{1}{2},\frac{1}{2},1 ;-\frac{1}{c}\right),\label{energy}\\
 S(a,b,c)=&\frac{\sqrt{\lambda_t}}{2c^{3/2}~\textrm{cos}\alpha_0}\sqrt{1+c~\textrm{cos}^2\alpha_0}~{}_2F_1\left(\frac{1}{2},\frac{3}{2},2 ;-\frac{1}{c}\right),
 \label{spin}
 \end{align}
whereas the global and R-charge are given by
  \begin{align}
  J_{\phi_{1}}=&\frac{2\sqrt{\lambda_t}\sqrt{a}}{3\sqrt{3}\pi}\int_{0}^{\theta_{1m}}\frac{(1-2\textrm{sin}^4\frac{\theta'_1}{2}+2\textrm{sin}^2\frac{\theta'_1}{2})+(1-2\textrm{sin}^2\frac{\theta'_1}{2})\frac{b}{a}}{\sqrt{1+4a\textrm{sin},^4\frac{\theta'_1}{2}+4(b-a)\textrm{sin}^2\frac{\theta'_1}{2}}}{}d\theta'_1,
  \label{global1}\\
  J_{\psi}=&\frac{4\sqrt{\lambda_t}\sqrt{a}}{3\sqrt{3}\pi}\int_{0}^{\theta_{1m}}\frac{(1-2\textrm{sin}^2\frac{\theta'_1}{2}+\frac{b}{4a})}{\sqrt{1+4a\textrm{sin}^4\frac{\theta'_1}{2}+4(b-a)\textrm{sin}^2\frac{\theta'_1}{2}}}d\theta'_1.
  \label{r1}
  \end{align}
  Computing the integrals we get the following expressions in terms of elliptic and hypergeometric functions
  \begin{align}
  J_{\phi_{1}}=&\frac{\sqrt{\lambda_t}}{3\sqrt{3}}g\sqrt{a}\left[\left(1-\frac{b}{a}+\frac{a+b+\sqrt{(a-b)^2-a}}{\sqrt{a}}\right) {}_2F_1\left(\frac{1}{2},\frac{1}{2},1;k^2\right)\right.\nonumber\\
  &\left.-\frac{1}{2\sqrt{a}}\left(\frac{a+b+\sqrt{(a-b)^2-a}}{a-b-\sqrt{(a-b)^2-a}}\right) {}_2F_1\left(-\frac{1}{2},\frac{1}{2},1;k^2\right)\right.\nonumber\\
  &\left.+\frac{4b}{\pi a}\Pi(\alpha^2,k)-\frac{2b}{\pi\sqrt{a}}\Pi(\beta^2,k)\right],\label{chargephi}\\
  J_\psi=&\frac{\sqrt{\lambda_t}}{3\sqrt{3}}g\sqrt{a}\left[ \left(\frac{b}{2a}-2\right)~{}_2F_1\left(\frac{1}{2},\frac{1}{2},1;k^2\right) +\frac{8}{\pi}\Pi(\alpha^2,k)\right],\label{chargepsi}
  \end{align}
  where 
  \begin{align}
  \alpha^2=&-\frac{1}{-1+2a-2b+2\sqrt{(a-b)^2-a}},\nonumber\\
  \beta^2=&-\frac{a-b-\sqrt{(a-b)^2-a}}{a+b+\sqrt{(a-b)^2-a}}.
  \end{align}
  Thus the expressions in eqs. (\ref{alphazero}), (\ref{energy}), (\ref{spin}), (\ref{chargephi}) and (\ref{chargepsi}) determine the energy in terms of the spin and the two angular momenta by eliminating the constants $a$, $b$ and $c$.   It is useful to study this functional dependence in some interesting limits. We will always be considering $J_T\gg1, ~S\gg1$ with $\sqrt{\lambda_t}\gg1$ such that $J_{T}/\sqrt{\lambda_t}$ and $S/\sqrt{\lambda_t}$ remain finite in the limit. 
  We
  have defined $J_{T}=(J_{\phi_1}+J_{\psi})$ as the total angular momentum of the string. According to the value of $S$ and $J_T$ (as compared to $\lambda_t$) we can also consider different cases. Of particular interest are the short and large string limits. The first case corresponds to small $S$ and $J_T$ and the second one to large $S$ and $J_T$ (as compared to $\lambda_t$). 
  
  We see from (\ref{spin}) that the limit of large and small $S$ corresponds, respectively, to $c\ll 1$ and $c\gg 1$. 
    One can also see from (\ref{chargephi}) and (\ref{chargepsi}) that the large total angular momentum limit corresponds to the parameter region where 
    $a\approx \frac{1+2b+\sqrt{1+4b}}{2}$ for fixed $b$, which implies $k^2\approx 1$, whilst the small region corresponds to $a\gg b$ where $k^2\ll1$.
  We now study these different limits.
 
  \textbf{Short strings}. In this limit (\ref{alphazero}) gives $\tan^2\alpha_0\approx\frac{J_{T}^2}{2S}$. Then (\ref{energy}) reduces to
  \begin{gather}
  E^2\cong J_{T}^2+2\sqrt{\lambda_t}S.
  \label{shortt11}
  \end{gather}
  This is a Regge type spectrum which corresponds to a string in flat space with total angular momentum
  $J_T$ rotating around its center of mass with spin $S$.
  
  \textbf{Large strings.} In this limit $\tan^{2}\alpha_0\approx\frac{ag^2\log\frac{16}{1-k^2}}{3\log^2 c}$. The expressions (\ref{energy}), (\ref{spin}), (\ref{chargephi}) and (\ref{chargepsi}) give
 \begin{gather}
 E-S\cong\frac{\sqrt{\lambda_t}}{\pi}\sqrt{\log^2 c+\frac{9\pi^2}{\mu_0^2}\left(\frac{ J_T}{\sqrt{\lambda_t}}-\frac{l_0}{3\sqrt{3}\pi}\right)^2},
  \end{gather}
 where $l_0=(1+4b)^{1/4}$ and $3/2<\mu_0<11/2$.
  So that we have reduced the problem to find $c$ in terms of $S$ and $J_T$ as a solution of the parametric equation
 \begin{gather}
 S=\frac{2\sqrt{\lambda_t}}{\pi c\vert\log c\vert}\sqrt{\log^2 c+\frac{9\pi^2}{\mu_0^2}\left(\frac{ J_T}{\sqrt{\lambda_t}}-\frac{l_0}{3\sqrt{3}\pi}\right)^2}.
 \label{parametric}
 \end{gather}
 When $J_T$ is small,  $\frac{J_T}{\sqrt{\lambda_t}}\ll \log c$, we get $c\approx\frac{2\sqrt{\lambda_t}}{\pi S}$. In this limit we find
 \begin{gather}
 E-S\cong\frac{\sqrt{\lambda_t}}{\pi}\log \frac{\pi S}{2\sqrt{\lambda_t}}+\frac{9\pi}{2\mu_0^2\sqrt{\lambda_t} }\frac{J_T^2}{\log \frac{\pi S}{2\sqrt{\lambda_t}}}.
 \label{larget11}
 \end{gather}
 which is the usual energy-spin dispersion relation for large spinning strings.
 
 In the fast string expansion, $\frac{J_T}{\sqrt{\lambda_t}}\gg \log c$, we have $c\log \frac{1}{c}\approx\frac{6 J_T}{\mu_0 S}$. Because $c\ll 1$ we have 
 $J_T\ll S$, thus the string is spinning faster in $AdS_5$. The leading order solution for $c$ is $c\approx  \frac{6J_T}{ \mu_0S}$,
we get
\begin{gather}
E-S\cong \frac{3}{\mu_0}J_T+\frac{\lambda_t\mu_0}{6\pi^2J_T}\log^2\frac{6J_T}{\mu_0S}+\frac{\sqrt{\lambda_t}l_0}{\sqrt{3}\mu_0\pi}.
\label{problematic1}
\end{gather}

Another interesting limit to consider, apart from the ones discussed above, is when \\
$\frac{J_T}{\sqrt{\lambda_t}}\gg 1$ and $\frac{S}{\sqrt{\lambda_t}}\ll1$ with $S/J_T$ fixed. In this case $\tan^{2}\alpha_0\approx a c g^2\log^2\frac{16}{1-k^2}$, 
and one finds that
\begin{gather}
E\cong \frac{3}{\mu_0}J_{T}+S+\frac{\sqrt{\lambda_t}S}{3\mu_0J_{T}}+\frac{\mu_0^2\lambda_t  S}{18J_T^2}+\frac{\sqrt{\lambda_t}l_0}{\sqrt{3}\pi\mu_0}.
\label{problematic2}
\end{gather}
This limit describes a near point-like string staying near the center of $AdS_5$.
We see that the first terms in (\ref{problematic2}) looks like the expansion in positive series of $\lambda_t$ which would suggest 
the direct comparison with field theory anomalous dimension of operators, however
the last term, being large in the limit, makes this comparison problematic.
\\
\\
For the sake of completeness,  let us discuss some particular cases of eq. (\ref{solution}) concerning different string
configurations in $T^{1,1}$.
 
 \subsubsection{Case $v_{\psi}=0$.}
The energy and spin are still given by eqs. (\ref{energy}) and (\ref{spin}), whereas the corresponding global charge is 
\begin{align}
  J_{\phi_{1}}=\frac{\sqrt{\lambda_t}}{3\sqrt{3}}\left(\frac{1}{4a}~{}_2F_1\left(\frac{1}{2},\frac{3}{2},2;\frac{1}{a} \right)+~{}_2F_1\left(\frac{1}{2},\frac{1}{2},1;\frac{1}{a}\right)\right).
  \label{chargesp}
\end{align}
Note that this charge is no longer equivalent to the one which is rotating only in the $S^2$ of $S^5$ for the maximal supersymmetric case \cite{Russo:2002sr,Frolov:2002av} due to
 the $U(1)$ fibration in $T^{1,1}$. Strictly speaking, the comparison is not quite direct because the scale in (\ref{chargesp}) is set by $T^{1,1}$ instead of
 $S^5$.
 
 We have different cases concerning the value of $a$. 
 \begin{itemize}
 \item For $a<1$,  then $\theta_1\in [0,\pi]$. The string is closed and stretched around the great circle of $S_1^2$ in $T^{1,1}$.
\item  For $a=1$, we have an infinite energy solution unless we fix $\theta_1$ and eq. (\ref{eqtheta}) implies that $\theta_1=\frac{\pi}{2}$.
\item For $a>1$ the string is stretched up to a maximal value determined by $\theta_{1m}=\arcsin \frac{1}{\sqrt{a}}$. 
This is the most interesting case and we shall study it in detail.
\end{itemize}
Consider the case $a>1$. We see from (\ref{chargesp}) that the regime of large and small global charge corresponds, respectively,  to the region where $a\approx 1$
and $a\ll 1$.

\textbf{Short strings.} In this case we find that $\tan^2\alpha_0\approx \frac{c\sqrt{3}J_{\phi_1}}{\sqrt{\lambda_t}}$.
The energy (\ref{energy}) becomes
\begin{gather}
E^{2}\approx\sqrt{\lambda_t}(2S+\sqrt{3}J_{\phi_1}),
\end{gather}
which is a Regge-type spectrum. 

 \textbf{Large strings.}
In this limit $\tan^2\alpha_0\approx\frac{6(\frac{\pi J_{\phi_1}}{\sqrt{\lambda_t}}+\frac{1}{3\sqrt{3}})^2}{\log^2c}$. 
We obtain
\begin{gather}
E-S\cong \sqrt{6\left(J_{\phi_1}+\frac{\sqrt{\lambda_t}}{3\sqrt{3}\pi}\right)^2+\frac{\lambda_t}{\pi^2}\log^2c}.
\end{gather}
For $\textrm{log}c\gg \frac{J_{\phi_1}}{\sqrt{\lambda_t}}$ we have $c\approx \frac{2\sqrt{\lambda_t}}{\pi S}$. Thus the energy-spin relation behaves like
\begin{gather}
E-S\cong\frac{\sqrt{\lambda_t}}{\pi}\textrm{log}(\frac{\pi S}{2\sqrt{\lambda_t}})+\frac{10\pi}{6\sqrt{\lambda_t}}\frac{J_{\phi_1}^2}{\log(\frac{\pi S}{2\sqrt{\lambda_t}})}.   \label{e1}
\end{gather}
For $\textrm{log}c\ll \frac{J_{\phi_1}}{\sqrt{\lambda_t}}$ we find $c\approx\frac{2\sqrt{6}J_{\phi_1}}{S}$, so that 
\begin{gather}
E-S\cong\sqrt{6}J_{\phi_1}+\frac{\lambda_t}{2\sqrt{6}\pi^2}\frac{\log\frac{2\sqrt{6}J_{\phi_1}}{S}}{J_{\phi_1}}+\frac{\sqrt{2}\sqrt{\lambda_t}}{3\pi}.
\label{e2}
\end{gather}
 This result agrees with
the result found in \cite{Schvellinger:2003vz}. Although, the relation in eq. (\ref{e2}) is the generalization to the non-zero spin case. 

\subsubsection{Case $v_{\phi_1}=0$.}
 In order to satisfy the periodicity condition we must set $\theta_1=\textrm{const}$, but eq. (\ref{eqtheta}) implies that $\theta_1=0$.
 We find that the R-charge is homogeneously distributed along the string 
\begin{gather}
\frac{J_{\psi}}{\sqrt{\lambda_t}}=\frac{v_{\psi}}{9}.
\end{gather}
An interesting limit to consider here is when $\frac{S}{\sqrt{\lambda_t}}\ll 1$ and $\frac{J_{\psi}}{\sqrt{\lambda_t}}\gg1$.
Then, the string is moving very fast along the R-charge direction with energy
\begin{gather}
E\cong 3J_{\psi}+S+\frac{\sqrt{\lambda_t}}{3}\frac{S}{J_{\psi}}+\frac{\lambda_t  S}{18 J_{\psi}^2}.
\label{mode}
\end{gather}
This energy is also captured by the leading quantum term in the spectrum of strings of (\ref{background})-(\ref{f5}), in the frame boosted to the speed of light along $\psi$.
In the strict $S=0$ limit we see that (\ref{mode}) saturates the BPS bound and  is dual to a chiral primary operator \cite{Gomis:2002km}.

As we will see in the next section, the subsector of states discussed above, will be preserved under dualisation
by imposing some conditions on the T-dual coordinates.

  \subsection{Classical solution for rotating strings in the Non-Abelian T-dual Klebanov-Witten Background}\label{sec3.2}
  Let us consider now a string moving in the background (\ref{penrose})-(\ref{dualfluxes}) which is rotating in the $\phi_1$ and the R-charge direction, 
  $\psi$, it is also stretched along $\theta_1$
  and localized at a fixed point in the plane $(x_1,x_2)$. From now on the string configuration for $AdS_5$ will be the one defined in eq. (\ref{solution}). 
  An appropriate ansatz for such a solution is
\begin{gather}
\theta_1=\theta_1(\sigma)=\theta_1(\sigma+2\pi), \qquad \phi_1=v_{\phi_1}\tau, \qquad \psi=v_{\psi}\tau,\nonumber\\
x_1=\textrm{fixed},\qquad x_{2}=\textrm{fixed}.
\label{tdualsol3}
\end{gather}
This configuration generates a subgroup of isometries $U(1)_1\times U(1)_R$ of (\ref{metricdual}) that has been retained by
non-Abelian T-duality. 
 The most general solution for  $\theta_1$ satisfying the conformal constraints is
 \begin{align}
 \theta'^{2}_1=&\left(4-4\frac{\lambda_2^2\lambda^2x_1^2}{\lambda^2_1\Delta}\right)v_{\phi_{1}}^2\textrm{sin}^4\frac{\theta_1}{2}+\left(\frac{4\lambda_2^2\lambda^2x_1^2}{\lambda_1^2\Delta}v_{\phi_{1}}v_{\psi}-\left(4-4\frac{\lambda_2^2\lambda^2x_1^2}{\lambda^2_1\Delta}\right)v_{\phi_{1}}^2\right)\textrm{sin}^2\frac{\theta_1}{2}\nonumber\\
 &+\frac{\kappa^2}{\lambda_1^2}\textrm{sin}^2\alpha_0-\frac{\lambda_2^2\lambda^2x_1^2}{\lambda_1^2\Delta}v_{\phi_1,\psi}^2,
  \label{soltdual}
 \end{align}
   where $v_{\phi_1,\psi}=v_{\phi_1}+v_{\psi}$ and $\Delta$ was defined in eq. (\ref{delta}). If we identify the analog to $a$ and $b$ in (\ref{parameters}) as
 \begin{align}
 a\rightarrow\hat{a}=&\frac{\left(4-4\frac{\lambda_2^2\lambda^2x_1^2}{\lambda^2_1\Delta}\right)v_{\phi_1}^2}{\frac{\kappa^2}{\lambda_1^2}\textrm{sin}^2\alpha_0-\frac{\lambda_2^2\lambda^2x_1^2}{\lambda_1^2\Delta}v_{\phi_1,\psi}^2},\nonumber\\
 b\rightarrow\hat{b}=&\frac{\frac{4\lambda_2^2\lambda^2x_1^2}{\lambda_1^2\Delta}v_{\phi_1}v_{\psi}}{\frac{\kappa^2}{\lambda_1^2}\textrm{sin}^2\alpha_0-\frac{\lambda_2^2\lambda^2x_1^2}{\lambda_1^2\Delta}v_{\phi_1,\psi}^2},
  \end{align}
  we can see that the expression in eq. (\ref{soltdual}) is equivalent to eq. (\ref{soltheta}). Thus, it seems that if we appropriately choose the values of $x_1$ and $x_2$
  we can get a set of states that remains unchanged under non-Abelian T-duality. This is indeed possible in a finite range determined by $ 1/3\sqrt{6}\leq x_1<1$ for any fixed $x_2$.
 The metric in eq. (\ref{metricdual}) around these values of $(x_1,x_2)$ behaves like
  \begin{gather}
  d\hat{s'}^2=\lambda_{\hat{1}}^{2}(\sigma_{\hat{1}}^2+\sigma_{\hat{2}}^2)+\frac{\lambda_2^2}{x_2^2+\lambda_2^4}(dx_1^2+x_1^2\sigma_{\hat{3}}^2).
  \label{tdualapprox}
  \end{gather}
  This metric has a bolt singularity \cite{Itsios:2012zv}. To remove this singularity
 one requires the range of $\psi$ to be $2\pi$. Then we see that the string states in (\ref{solution}) and the ones in (\ref{tdualsol3}) 
 will be characterized by the same labels (\ref{energy}), (\ref{spin}), 
 (\ref{chargephi}) and (\ref{chargepsi}) up to a $\mathbb{Z}_2$ quotient on $\psi$.
 
 The latter result implies that, even though the field theory dual to the background in eqs .(\ref{penrose})-(\ref{dualfluxes}) is, in principle, different to the field theory
 dual to the background in eqs. (\ref{background})-(\ref{f5}), there is a particular sector where both are equivalent. This is due to the fact that a precise sector of the
 geometry, apart from $AdS_5$, was unaffected by non-Abelian T-duality. In other words, the computation of a given observable that in the original background 
 is {\it uncharged} under $SU(2)_2$ will give the same result in the T-dual background. One may thus expect that the computation of
 observables in that subsector will go through. However, we have seen that this is not the case; the equivalence is guaranteed up to certain conditions on the T-dual coordinates.
 This unexpected behavior is because the term containing the R-charge direction in eq. (\ref{metricdual}), has been mixed non-trivially by T-duality with the T-dual coordinates. In 
 addition, note that the fact that the R-charge direction was retained
 and made explicit in the T-dual solution is a consequence of the gauge choice used in Sec. \ref{sec2};  we can still identify it as the R-charge direction 
 but by no means is the same.

 We shall study now configurations where the string is stretched along the T-dual coordinates. 
    
A configuration defining a string which is rotating along the $\phi_1$ and the R-charge direction on the equator of $S_1^3$ in (\ref{metricdual}) 
that is located at  some fixed point $x_2$ and stretched along $x_1(\sigma)$, can be parametrized by the ansatz
\begin{gather}
\theta_1=\frac{\pi}{2}, \qquad \phi_1=v_{\phi1}\tau, \qquad \psi=v_{\psi}\tau,\nonumber\\
x_1(\sigma)=x_1(\sigma+2\pi),\qquad x_{2}=\textrm{fixed}.
\label{tdualsol1}
\end{gather}
The equation of motion for $x_1$ becomes
\begin{gather}
(x_1^2+\lambda_2^2\lambda^2)x''_1+\frac{\lambda^2 x_2^2 x_1(x'_{1})^2}{\Delta}+\frac{\lambda^2\lambda_2^2x_1(x_2^2+\lambda_2^4)}{\Delta}v_{\psi}^2=0.
\label{eqx1}
\end{gather}
The off-diagonal components of the conformal constraints implies different cases.
\begin{itemize}
\item For $v_{\psi}\neq$0, a general solution of eq. (\ref{eqx1}) is
  \begin{gather}
(x_1^2+\lambda^2\lambda_2^2)(x'_{1})^2=\lambda^2_2(\kappa^2\sin^2\alpha_0-\lambda_1^2v_{\phi_1}^2)\left(\lambda_2^2\lambda^2-(l-1)x_1^2\right),\\
 x_2=0,  
 \label{eqx2}
  \end{gather}
where
  \begin{gather}
  l=\frac{\lambda^2v_{\psi}^2}{(\kappa^2\sin^2\alpha_0-\lambda_1^2v_{\phi_1}^2)}.
  \label{paramdual}
  \end{gather}
  For $l>1$, $x_1$ has a maximum $x_{1m}$, defined by the equation $x'_1=0$. 
  As in the previous section, for the single-fold case, $x_1$ starts in $x_1=0$ at $\sigma=0$
  and increases up to reach its maximal value $x_{1m}$ at $\sigma=\pi/2$. By demanding this we get the condition
 \begin{align}
 \tan^{2}\alpha_0=&\frac{\left(\frac{\lambda^2 l}{(l-1)^2}~{}_2F^2_{1}\left(-\frac{1}{2},\frac{1}{2},1 ;\frac{1}{l}\right)+\lambda_1^2v_{\phi_1}^2\right)c}{{}_2F^2_{1}\left(\frac{1}{2},\frac{1}{2},1; \frac{1}{l}\right)}\label{periodt}.
 \end{align}
 The global charge and R-charge for this soliton are 
  \begin{align}
  J_{\phi_{1}}=&\sqrt{\lambda_t}\lambda_1^{2}v_{\phi_1},\label{gchargedual}\\
 J_{\psi} =&\frac{\sqrt{\lambda_t}l}{l-1}\left[{}_2 F_{1}\left(-\frac{1}{2},\frac{1}{2},1; \frac{1}{l}\right)-\frac{(l-1)}{l}{}_2 F_{1}\left(\frac{1}{2},\frac{1}{2},1 ;\frac{1}{l}\right)\right].
 \label{rcharget}
  \end{align}

  Whereas the energy and spin are still given  by eqs. (\ref{energy}) and (\ref{spin}). 
    
  Note that the homogeneous distribution of angular momentum (\ref{gchargedual}) is bounded in virtue of (\ref{paramdual}).
  We also see that the large R-charge limit is not allowed for this configuration. Therefore, we will have only short strings.
   In such case $l\gg1$, we see that
  $\kappa\sim v_{\phi_1}\approx0$.
  
\textbf{Short strings} In this limit the energy (\ref{energy}) behaves like
  \begin{gather}
  E^{2}\cong\sqrt{\lambda_t}\left(2S+J_{\psi}\right)+6J_{\phi_1}^2,
  \label{shortt}
  \end{gather}
 which is the usual Regge behavior with a small correction due to $J_{\phi_1}$. 
 
\item For $v_{\psi}=0$,
the periodicity condition on $x_1$ implies that $x_1=\textrm{const}$ and eq. (\ref{eqx1}) sets it to zero.
 Thus, the vanishing of the R-charge implies the vanishing of the T-dual coordinate $x_1$.
 \end{itemize}
    
    Let us consider now that the string is stretched along the T-dual coordinate $x_2$ and we seek now for a solution as in (\ref{tdualsol1})
   but now with $x_1=\textrm{fixed},~ x_2=x_2(\sigma)=x_2(\sigma+2\pi)$.
The equation of motion for $x_2(\sigma)$ becomes
\begin{gather}
(x_2^2+\lambda_2^2)x''_2+\frac{x_2x_1^2\lambda_2^2}{\Delta}((x'_2)^2-\lambda^4v_{\psi}^2)=0.
\label{condx2}
\end{gather}
The conformal constraints imply that $x_2=\textrm{fixed}$  and eq. (\ref{condx2}) sets it to zero. 
 In this case the global and R-charge are uniformly distributed along the string
\begin{gather}
\frac{J_{\phi_1}}{\sqrt{\lambda_t}}=v_{\phi_{1}},\qquad  \frac{J_{\psi}}{\sqrt{\lambda_t}}=\frac{\lambda^2 x_1^2}{x_1^2+\lambda^2\lambda_2^2}v_{\psi}.
\end{gather}
We then see that $x_1=0$ implies zero R-charge. 
 Let us consider the case of non-vanishing $x_1$ and $v_{\phi_1}=0$ with $\frac{S}{\sqrt{\lambda_t}}\ll1$, $\frac{J_{\psi}}{\sqrt{\lambda_t}}\gg 1$. 
 In this limit the energy is
\begin{gather}
E\cong \frac{3\sqrt{x_1^2+\lambda_1^2\lambda^2}}{x_1}J_{\psi}+S+\frac{\sqrt{\lambda_t}x_1S}{3\sqrt{x_1^2+\lambda_1^2\lambda^2}J_{\psi}}+\frac{\lambda_t x_1^2S}{18(x_1^2+\lambda_1^2\lambda^2)J^2_{\psi}}.
\label{dualbps}
\end{gather}
In the limit $S=0$, the point-like string is placed at the center of $AdS_5$ and is rotating along the R-charge direction with the speed of light.
This is a prediction for the dual (strongly coupled) field theory; there will be a set of chiral primary operators with a scaled R-charge due to $x_1$. 
In other words, the anomalous dimension of these operators in the dual field theory will depend on the (fixed) T-dual coordinate $x_1$.
We see from (\ref{dualbps}) that the point $x_1=0$ will be problematic. One might try to chose $x_1=0$ by considering $J_{\psi}/x_1$ fixed, but this is indeed not possible
in the present  case.

The results of this section implies that we cannot have a description of string states with large quantum numbers
independent of the value of the T-dual coordinates. Particularly, we have analyzed how the states discussed in Sec. \ref{sec3.1} 
has changed. We found that  for specific values of the T-dual coordinates such states become equivalent. 
Moreover, some states are not only affected but their existence is constrained by the T-dual coordinates. For instance, there is a correlation between the values of 
$x_1$ and $\psi$ such that a necessary, but however not sufficient, condition to have non-zero R-charge is not to set $x_1$ to zero.
Note that for all the solutions discussed above the NS field in eq. (\ref{bfield}) does not introduce corrections.


 \section{Towards the Penrose limit of the dualised background}\label{sec5}
A remarkable observation made in \cite{Metsaev:2001bj} is that type IIB superstring theory on the pp-wave background with a five form RR flux is exactly solvable.
In addition, this background is also a maximally supersymmetric solution of type IIB supergravity \cite{Blau:2001ne} which 
can be obtained as the Penrose limit of another maximal supersymmetric solution, $AdS_5\times S^5$. These ideas were first studied within the AdS/CFT 
correspondence in \cite{Berenstein:2002jq} to identify stringy states in $AdS_5\times S^5$  with a class of gauge invariant operators of the $\mathcal{N}=4$ $SU(N)$ 
supersymmetric gauge theory. Since then, the analysis of pp-wave backgrounds coming from the Penrose limit of supergravity 
backgrounds has been extended to a number of models (see, e.g. \cite{Sadri:2003pr}). 

Interestingly, in \cite{Gueven:2000ru}  it was shown that the scaling limit is consistent with T-duality. The consistence relies in proving that the gauge choices
for the NS and RR fields are in harmony with the T-duality rules for the corresponding transformed fields such that they produce 
finite non-zero results in the limit.
Therefore, the T-dual pp-waves are also solutions and consequently this limiting procedure commutes with T-duality. This idea has been applied to type 
IIB backgrounds to generate new pp-wave solutions of type IIA supergravity which preserve some fraction of supersymmetry  \cite{Michelson:2002wa}.
It is then natural to ask how the Penrose limit behaves under non-Abelian T-duality. 
However, in spite of the non-Abelian T-dual solution discussed here, and in general the ones obtained in \cite{Nunez}, has proven to be solution of type IIA
 supergravity, this transformation has led to topology changes that make not evident the applicability of such limiting procedure.
Nevertheless, the semiclassical analysis performed in Sec.  \ref{sec3.2} gives us strong evidence that such limit indeed exists. Then we ought to, in principle, reconsider 
the Penrose-Gueven derivation in a formalism suitable to incorporate the non-Abelian T-duality Rules for the NS and RR fields. 
We leave this study for a future work. For the time being we focus our attention on the metric
only and we shall show that we can indeed write it as a pp-wave metric.

 Let us consider a scaling limit around a null geodesic at a fixed point $x_1$ near $\rho=\theta_1=x_2=0$ in the background metric (\ref{penrose}) carrying
 large angular momentum along the $\psi$ direction.
 The coordinates that label the geodesic are 
 \begin{gather}
 x^{+}=\frac{1}{2}\left(t+\frac{x_{1}(\psi+\phi_1)}{3\sqrt{x_{1}^2+\lambda^2\lambda_2^2}}\right), \qquad x^{-}=\frac{L^2}{2}\left(t-\frac{x_{1}(\psi+\phi_1)}{3\sqrt{x_{1}^2+\lambda^2\lambda_2^2}}\right).
 \label{lightcone}
 \end{gather}
 We take $L\rightarrow \infty$ while rescaling the coordinates 
 \begin{gather}
 \rho=\frac{r}{L},\qquad \theta_1=\frac{\sqrt{6}}{L}\xi,\qquad x_2=\frac{\sqrt{x_{1}^2+\lambda^2\lambda_2^2}}{\lambda_2 L}\gamma .
 \end{gather}
 In this limit the metric becomes
 \begin{align}
 ds^2=&-4dx^+dx^-+\sum_{i=1}^{4}((dr^{i}dr^{i}+d\gamma^2)-(r^{i}r^{i}+4\gamma^2)dx^{+}dx^{+})\nonumber\\
 +&(d\xi_1^2+\xi_1^2d\phi_1^2)-2\frac{ x_1}{\sqrt{x_{1}^2+\lambda^2\lambda_2^2}}\xi_1^2dx^{+}d\phi_1.
 \label{metricpp}
 \end{align}
One easily sees that this is a pp-wave metric in virtue of $\partial_{-}$ being a covariantly constant null Killing vector.
This background has a number of isometries, some of which are manifest. In particular the transverse $SO(8)$ invariance of (\ref{metricpp})
 has been broken down to $SO(4)\times U(1)_{1}$. It can be seen that this pp-wave metric has less isometries that the 
 one obtained by applying the scaling limit to the background metric (\ref{background}) \cite{Gomis:2002km}. Thus we anticipate that the scaling limit
 does not commute with non-Abelian T-duality.
 
By considering the correspondence between conserved 
 charges in string theory and field theory one finds
 \begin{align}
2p^{-}=&i\partial_{x^{+}}=i\left(\partial_t+3\frac{\sqrt{x_1^2+\lambda^2\lambda_1^2}}{x_1}\partial_{\psi}\right)\nonumber \\
=&\Delta-\frac{3}{2}\frac{\sqrt{x_1^2+\lambda^2\lambda_1^2}}{x_1}R,\\
2p^{+}=&i\partial_{x^{-}}=\frac{i}{L^2}\left(\partial_t-3\frac{\sqrt{x_1^2+\lambda^2\lambda_1^2}}{x_1}\partial_{\psi}\right)\nonumber\\
=&\frac{1}{L^2}\left(\Delta+\frac{3}{2}\frac{\sqrt{x_1^2-\lambda^2\lambda_1^2}}{x_1}R\right).
\end{align}
Thus, as expected, this analysis predicts the existence of chiral primary operators whose anomalous dimension is 
$x_1$-dependent. This also predicts a set of operators whose deviation from the BPS bound is finite in the large $N$ limit.

 \section{Conclusions}\label{sec6}
 In this work we have discussed semiclassical solutions for rotating and spinning closed strings in the KW and the T-dual 
 KW backgrounds with two components of angular momentum which correspond to the global $U(1)_1$ and $U(1)_R$ charges
 of the internal space. Such configurations exhaust the number of isometries that we can have before and after the dualisation.
 
 We began by considering this configuration in the KW background by keeping fixed two of the coordinates of $SU(2)_2$
 about which we dualise the background. We found analytical expressions for the angular momenta.
 We then studied the behavior of the  string energy  in terms of these conserved quantities and the spin to recover the expected features for
  short and large strings as well as the holographic bound for the supergravity modes in this background.
We then studied this configuration in the T-dual background by keeping fixed the T-dual coordinates. We found that both backgrounds enjoy an 
equivalent subsector of states. However, the existence of this equivalent
sector depends on the values of the T-dual coordinates. The geometry displayed in this subsector probes to be a squashed 
three-sphere. This geometry induces also a reduction in the range of the R-charge direction in order to define a smooth equivalence.
We also studied configurations where the string is stretched along the T-dual coordinates. For a string that is stretched along $x_1$ for fixed 
$x_2$, we found that only short strings are allowed. In the case of vanishing R-charge we found that the string 
shrinks to a point. There is not a possible configuration where the string is stretched along  $x_2$ for fixed $x_1$. In this case we found that
the string is placed at a fixed point $(x_1,0)$ in the $(x_1,x_2)$ plane with a homogenous distribution of global and R-charge. The R-charge is $x_1$-dependent 
such that for vanishing $x_1$ the R-charge is identically zero. For non vanishing $x_1$, in the point-particle limit, the 
corresponding supergravity mode saturates the BPS bound but with a scaled R-charge related to the fixed value of $x_1$.
This puzzling feature  predicts that the anomalous dimension of operators in the dual superconformal field theory is $x_1$-dependent.
This discussion motivates that the T-dual background goes over to a pp-wave solution in the Penrose limit. 
We discussed briefly the application of this scaling limit to the T-dual fields. Particularly we worked with the metric by proving that we can 
indeed write it as a pp-wave metric. Then we wrote down the equivalence between conserved charges in string theory and field theory
which predicts the existence of the states we found by performing the semiclassical analysis.
This latter fact gives strong evidence for the generalization of this scaling
limit to the inclusion of the non-Abelian transformation for the other background fields which will be analyzed in a future work.

\section{Acknowledgements}
It is a pleasure to thank Carlos N\'u\~{n}ez for the many useful discussions during this project
and for help on improving this manuscript. I  would also thank to K. Sfetsos and J.A. Preciado for useful comments to the manuscript.
Thanks are also extended to the Physics Department of Swansea University for kind hospitality during the development of the present project.
This work was supported by a CONACYT-M\'exico Scholarship and UG PIFI projects.


\end{document}